\documentclass[runningheads]{llncs}
\usepackage{amsfonts}
\usepackage{amsmath}
\usepackage{graphicx}
\usepackage{mathtools}
\usepackage{bm}
\usepackage{siunitx}
\usepackage{hyperref}
\begin{document}
\title{Gradient-Based Geometry Learning for Fan-Beam CT Reconstruction}
\titlerunning{Gradient-Based CT Geometry Learning}
%

\author{Mareike Thies\inst{1} \and
Fabian Wagner\inst{1} \and
Noah Maul\inst{1,2} \and
Lukas Folle\inst{1} \and
Manuela Meier\inst{1,2} \and
Maximilian Rohleder\inst{1,2} \and
Linda-Sophie Schneider\inst{1} \and
Laura Pfaff\inst{1,2} \and
Mingxuan Gu\inst{1} \and
Jonas Utz\inst{3} \and
Felix Denzinger\inst{1,2} \and
Michael Manhart\inst{2} \and
Andreas Maier\inst{1}
}

\authorrunning{M. Thies et al.}

%

\institute{Pattern Recognition Lab, FAU Erlangen-N\"urnberg, Germany \and
Siemens Healthcare GmbH, Erlangen, Germany \and
Department AIBE, FAU Erlangen-N\"urnberg, Germany
}

\maketitle              
\begin{abstract}
Incorporating computed tomography (CT) reconstruction operators into differentiable pipelines has proven beneficial in many applications. Such approaches usually focus on the projection data and keep the acquisition geometry fixed. However, precise knowledge of the acquisition geometry is essential for high quality reconstruction results. In this paper, the differentiable formulation of fan-beam CT reconstruction is extended to the acquisition geometry. This allows to propagate gradient information from a loss function on the reconstructed image into the geometry parameters. As a proof-of-concept experiment, this idea is applied to rigid motion compensation. The cost function is parameterized by a trained neural network which regresses an image quality metric from the motion affected reconstruction alone. Using the proposed method, we are the first to optimize such an autofocus-inspired algorithm based on analytical gradients. The algorithm achieves a reduction in MSE by \SI{35.5}{\percent} and an improvement in SSIM by \SI{12.6}{\percent} over the motion affected reconstruction. Next to motion compensation, we see further use cases of our differentiable method for scanner calibration or hybrid techniques employing deep models.

\end{abstract}

\section{Introduction}
Artifact-free computed tomography (CT) reconstruction depends crucially on the exact knowledge of the acquisition geometry. It defines the rays on which X-rays penetrate the patient and the location at which the remaining X-ray photons are detected. Inaccurate models of the acquisition geometry in the reconstruction algorithm lead to artifacts such as blur or streaks in the reconstructed image. Patient motion falsifies the calibrated acquisition geometry because it alters the rays through the patient being measured. Hence, it is crucial to update the acquisition geometry assumed during reconstruction such that it compensates for global, rigid patient motion. One class of motion compensation approaches defines a quality metric (QM) on the reconstructed image, often referred to as autofocus criterion. This QM is minimized with respect to the geometry parameters to find an acquisition geometry which annihilates the patient motion and maximizes the quality of the reconstructed image \cite{capostagno2021,huang2022,kingston2011,preuhs2020,sisniega2017}. 
While the exact formulation of the QM and the parameterization of the motion may vary, all of these approaches have in common that they rely on gradient-free optimization. 

Moreover, differentiable CT reconstruction modules can be embedded into neural network architectures \cite{ronchetti2020,syben2019,wagner2022}. This is possible because they define the analytical gradients necessary for gradient-based optimization which drives all common deep learning frameworks. However, in all prior works, these analytical gradients are only computed with respect to the projection detector data. There exists no differentiable CT reconstruction operator which allows for gradient backpropagation into the geometry parameters. 

The two mentioned applications - autofocus motion compensation and neural network embedded CT reconstruction modules - are examples which can profit from derivatives of the reconstructed image with respect to the acquisition geometry. We hypothesize that the reason for which analytical geometry derivatives have not been incorporated into either of these problems is that computing and implementing these derivatives is not trivial. Previous work relies entirely on automatic differentiation for gradient-driven estimation of geometry parameters \cite{genzel2022,ruckert2022}. Such implementations prohibit strong parallelization on GPU and hence are not tractable for large problem sizes encountered in practice. Similarly, a cone-beam CT forward projection operator with differentiable geometry has been proposed as an extension to the idea of spatial transformers \cite{gao2020}. This technique is very memory intensive and not straight-forward to extend for backprojection. 

In this paper, fan-beam CT filtered backprojection is analytically derived with respect to the acquisition geometry. Precisely, we compute the Jacobian matrix which contains the derivative of each pixel in the reconstructed image with respect to each entry in the $2 \times 3$ projection matrices defining the acquisition geometry. We demonstrate the usefulness of our efforts by pairing the geometry-differentiable reconstruction with an autofocus objective and performing motion compensation via gradient descent. All implementations are readily integrated into the deep learning framework \textit{PyTorch} which facilitates optimization and highlights the potential of our implementations in the context of deep learning. The source code will be made available upon publication. To summarize, our contributions are:
\begin{itemize}
    \item The derivation of analytical gradients of a reconstructed image with respect to the projection matrices for fan-beam geometries in CT,
    \item the correct and GPU-accelerated implementation thereof as a differentiable operator for deep learning algorithms, and
    \item proof-of-concept experiments highlighting the value of our method for CT motion compensation by translation of previous work on gradient-free autofocus motion compensation approaches into the gradient-based setting.
\end{itemize}

\begin{figure}
    \centering
    \includegraphics[trim={5cm 0cm 0cm 0cm},clip,width=\textwidth]{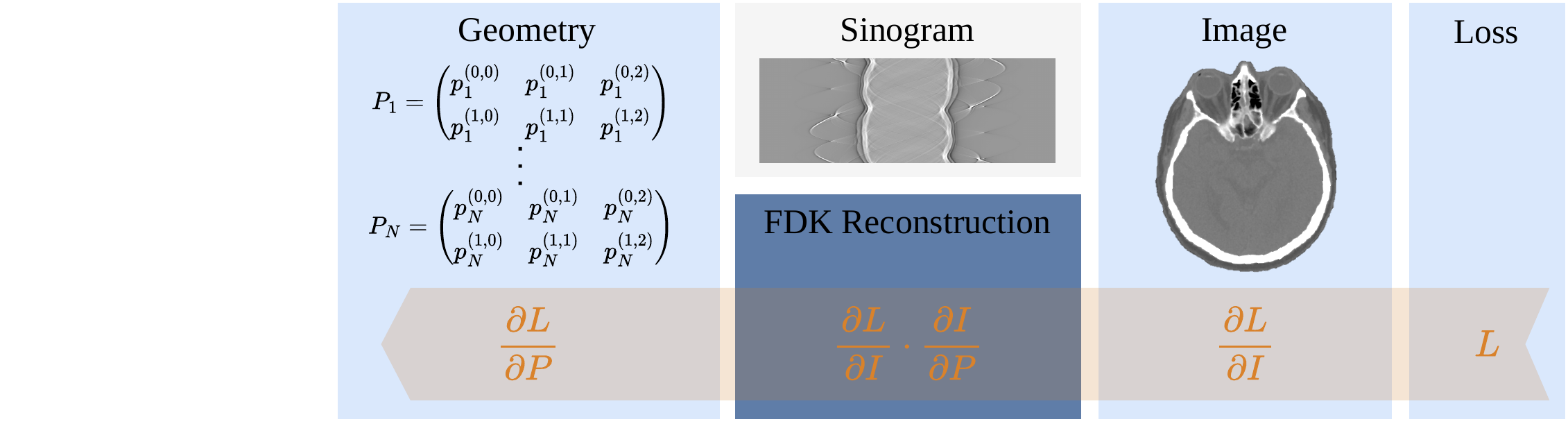}
    \caption{We propose a method to propagate the gradient of a loss $L$ with respect to a reconstructed image into the projection matrices parameterizing the fan-beam geometry. The crucial step is the computation of the Jacobian matrix $\frac{\partial I}{\partial P}$ which contains the partial derivatives of the gray values in the reconstructed image $I$ with respect to the entries of the projection matrices $P$.}  
    \label{fig:my_label}
\end{figure}

\section{Methods}
\subsection{Derivation of Analytical Gradients}
\label{sec:gradients}
\subsubsection{Projection Matrices}
A widely used parameterization of CT acquisition geometry is the projection matrix. For the fan-beam case studied here, the geometry of each projection is represented by a projection matrix $\bm{P}_i \in \mathbb{R}^{2 \times 3}$, $i=1,...,N$ with $N$ being the number of projections in a full scan. Each matrix $\bm{P}_i$ consists of an extrinsic part $\bm{E}_i \in \mathbb{R}^{2 \times 3}$ and an intrinsic part $\bm{K}_i \in \mathbb{R}^{2 \times 2}$. The extrinsic matrix models rotation and translation of the imaged object while the intrinsic matrix describes parameters of the detector, such as the detector pixel size and the detector origin. The product of intrinsic and extrinsic matrix yields the projection matrix
\begin{equation}
    \bm{P}_i = \bm{K}_i \cdot \bm{E}_i = \bm{K}_i \cdot [\bm{R}_i | \bm{t}_i] \enspace,
\end{equation}
with $\bm{R}_i \in \mathbb{R}^{2\times 2}$ being a 2D rotation matrix and $\bm{t}_i \in \mathbb{R}^{2}$ being a translation vector. 
Using homogeneous coordinates, the non-linear perspective mapping inherent to fan-beam geometries can be expressed as a linear mapping
\begin{equation}
    \label{eq:2}
    \begin{pmatrix} 
        u_i\\
        v_i
    \end{pmatrix} 
    = \bm{P}_i \cdot
    \begin{pmatrix} 
        x\\
        y\\
        1
    \end{pmatrix} 
     = 
    \begin{pmatrix} 
        p_i^{(0,0)}x + p_i^{(0,1)}y + p_i^{(0,2)}\\
        p_i^{(1,0)}x + p_i^{(1,1)}y + p_i^{(1,2)}
    \end{pmatrix} 
    \enspace .
\end{equation}
We follow the notation in \cite{aichert2015} and denote the space of all homogeneous representations of $n$-dimensional vectors as $\mathbb{P}^{n+}$. Then, the homogeneous image point $(x, y, 1)^T \in \mathbb{P}^{2+}$ gets mapped onto the homogeneous detector position $(u_i, v_i) \in \mathbb{P}^{1+}$ via multiplication with the projection matrix $\bm{P}_i$. We further define $g: \mathbb{P}^{1+} \rightarrow \mathbb{R}$ as the function which restores the Euclidean coordinate from the homogeneous representation by dividing the vector by its last component
\begin{equation}
    \label{eq:3}
    g \left( (u_i, v_i)^T \right) = \frac{u_i}{v_i} \eqqcolon w_i \enspace ,
\end{equation}
where $w_i \in \mathbb{R}$ is the 1D position on the detector onto which the image point $(x, y)^T$ gets mapped under projection with projection matrix $\bm{P}_i$.

\subsubsection{Fan-Beam Reconstruction}
The value of a position $(x, y)^T$ in the reconstructed image $\bm{I}$ is
\begin{equation}
    \label{eq:4}
    \bm{I}(x, y) = \sum_{i=1}^{N} d_i \left(w_i \right) = \sum_{i=1}^{N} d_i \left( g \left( \bm{P}_i \cdot
    \begin{pmatrix} 
        x\\
        y\\
        1
    \end{pmatrix} 
    \right) \right) \enspace .
\end{equation}
The position $(x, y)^T$ in the reconstructed image is forward projected into each projection $i$ using Eq.~\ref{eq:2} and \ref{eq:3}. The projection data consists of ramp filtered detector signals $\bm{D}_i \in \mathbb{R}^M$, $i = 1,...,N$ where $M$ is the number of detector pixels. The function $d_i:\mathbb{R} \rightarrow \mathbb{R}$ interpolates the detector signal $\bm{D}_i$ at the forward projected position $w_i$. The interpolated values are summed up for all views $i$ to obtain the reconstructed value at position $(x, y)^T$. A full reconstructed image $\bm{I}$ is computed by evaluating Eq.~\ref{eq:4} on a discrete 2D grid in $x$ and $y$.

\subsubsection{Gradient Computations}
We derive an expression for the derivative of a single reconstructed gray value in the image $\bm{I}$ at position $(x,y)^T$ with respect to all 6 entries of the $j$-th projection matrix
\begin{equation}
    \label{eq:5}
    \frac{\partial \bm{I}(x,y)}{\partial \bm{P}_j} =  \frac{\partial}{\partial \bm{P}_j} \left\{ d_j \left( g \left( \bm{P}_j \cdot
    \begin{pmatrix} 
        x\\
        y\\
        1
    \end{pmatrix} 
    \right) \right) \right\} \enspace .
\end{equation}
The sum in Eq.~\ref{eq:4} disappears because only the $j$-th summand depends on $P_j$. As Eq.~\ref{eq:5} is composed of three nested functions, we apply the chain rule of differentiation to break the computation down into three simpler derivatives  
\begin{equation}
    \label{eq:6}
    \frac{\partial \bm{I}(x,y)}{\partial \bm{P}_j} = \frac{\partial d_j}{\partial w} \bigg\rvert_{w=w_j} \cdot \frac{\partial w}{\partial (u, v)^T} \bigg\rvert_{\substack{u=u_j \\ v=v_j}} \cdot \frac{\partial (u,v)^T}{\partial \bm{P}} \bigg\rvert_{\bm{P}=\bm{P}_j} \enspace .
\end{equation}
The first chained gradient describes the derivative of the interpolated $j$-th ramp filtered detector signal with respect to the interpolation position. This can be computed using a finite differences approximation of the gradient on the gray values of the ramp filtered detector signal along the detector elements followed by interpolation at position $w_j$. Given the prefiltered detector signal $\bm{D}^{*}_j \in \mathbb{R}^M$, function $d^*_j: \mathbb{R} \rightarrow \mathbb{R}$ performs the interpolation of $\bm{D}_j^*$ at position $w_j$ . The second and third derivatives follow directly from Eq.~\ref{eq:3} and \ref{eq:2} (linearized), respectively. 
\begin{align}
    \label{eq:7}
    \frac{\partial d_j}{\partial w} & \bigg\rvert_{w=w_j} = d^*_j(w_j) &&\in \mathbb{R} \\
    \label{eq:8}
    \frac{\partial w}{\partial (u, v)^T} & \bigg\rvert_{\substack{u=u_j \\ v=v_j}} \, = \left( \frac{1}{v_j}, -\frac{u_j}{v_j^2} \right) &&\in \mathbb{R}^{1 \times 2} \\
    \label{eq:9}
    \frac{\partial (u,v)^T}{\partial \bm{P}} & \bigg\rvert_{\bm{P}=\bm{P_j}} = \begin{pmatrix} 
        x\ & y\ & 1\ & 0\ & 0\ & 0\ \\
        0 & 0 & 0 & x & y & 1
    \end{pmatrix} &&\in \mathbb{R}^{2 \times 6}
\end{align}
Inserting Eq.~\ref{eq:7}--\ref{eq:9} into Eq.~\ref{eq:6} yields the final $1 \times 6$ dimensional gradient of the reconstructed gray value in image $\bm{I}$ at position $(x,y)^T$ with respect to the six entries of projection matrix $\bm{P}_j$. Given a full reconstructed image $\bm{I}$, this computation needs to be carried out for each pixel position in the image and for each projection matrix $\bm{P}_i$, $i=1,...,N$. This yields a full Jacobian matrix $\frac{\partial \bm{I}}{\partial \bm{P}}$. 

\subsubsection{Implementation}
We implement a geometry differentiable fan-beam reconstruction as a \textit{PyTorch} layer, i.e., subclass of \texttt{torch.autograd.Function}. This way we can utilize the built-in optimization functionality of the \textit{PyTorch} framework. By specifying a loss function $L$ on the reconstructed image, its gradient with respect to the reconstructed image $\frac{\partial L}{\partial \bm{I}}$ can be obtained via automatic differentiation. Together with the analytical derivative $\frac{\partial \bm{I}}{\partial \bm{P}}$ calculated above, the gradient of the loss can be propagated into the acquisition geometry
\begin{equation}
    \frac{\partial L}{\partial \bm{P}} = \frac{\partial L}{\partial \bm{I}} \cdot \frac{\partial \bm{I}}{\partial \bm{P}} \enspace .
\end{equation}
The sampling positions $u_j$ and $v_j$ are required for the gradient computation in Eq.~\ref{eq:8}. We follow a dynamic programming approach and save the positions during the forward pass to avoid repetitive computations. The spatial derivative of the ramp filtered detector signal $\bm{D}_j^*$ needed in Eq.~\ref{eq:7} is computed once per backward call using second-order accurate central differences and is then evaluated multiple times at positions $w_j = u_j / v_j$. The functions $d_j$ and $d_j^*$ interpolate the (prefiltered) detector signal by performing simple linear interpolation along the detector elements. Both reconstruction and gradient computations are parallelized on GPU. 

\subsection{Motion Compensation}
We consider the problem of updating an initial set of projection matrices such that it compensates for random, inter-frame patient motion. All motion is assumed to be rigid, incorporating only rotation and translation in image space. 
Thus, the task is to find a matrix $\bm{M}_i \in \mathbb{R}^{3 \times 3}$, $i=1,..,N$ as a multiplicative geometry correction for each projection matrix
\begin{equation}
    \label{eq:11}
    \bm{P}^*_i = \bm{P}_i \cdot \bm{M}_i = \bm{P}_i \cdot \left(\begin{array}{c|c}
    \bm{R}(\alpha_i) \; & \; \bm{t}_i\\ \hline  
    \bm{0} \; & \; 1 
    \end{array}\right)
    = \bm{P}_i \cdot \begin{pmatrix}
    \cos{\alpha_i} \ & -\sin{\alpha_i} \ & t_{x,i}\\ 
    \sin{\alpha_i} \ & \cos{\alpha_i} \ & t_{y,i}\\ 
    0 \ & 0 \  & 1 
    \end{pmatrix} \enspace ,
\end{equation}
where each $\bm{M}_i$ is composed of a 2D rotation matrix $\bm{R}(\alpha_i) \in \mathbb{R}^{2 \times 2}$ and a translation vector $\bm{t}_i \in \mathbb{R}^2$. As each $\bm{R}(\alpha_i)$ depends on a single rotation angle $\alpha_i$, the dimensionality of this problem is $3N$. The free parameters are fitted by means of an optimization-based algorithm. Leveraging the autofocus idea, a target function is formulated on the reconstructed, motion-corrupted image itself. Typical choices are, e.g., image entropy, total variation, or gradient variance \cite{capostagno2021,wicklein2012}. These measures promote image sharpness or piece-wise constancy, but are agnostic to the underlying anatomy. To counteract convergence to anatomically implausible solutions, previous work suggests to train a deep neural network (DNN) to regress a certain QM from the motion corrupted images \cite{huang2022,preuhs2020,sisniega2021}. To demonstrate the advantages of the gradients derived in~\ref{sec:gradients}, we follow this idea and train a DNN-based autofocus target function $f_{\bm{\theta}}:\mathbb{R}^{N_x\times N_y}\rightarrow\mathbb{R}$ which regresses the structural similarity index measure (SSIM) between the motion-corrupted and motion-free reconstructed image given only the motion-corrupted image. The network is parameterized by network weights $\bm{\theta}$ and $N_x, N_y \in \mathbb{R}$ are the reconstructed image dimensions. Once trained, all existing autofocus approaches utilize the DNN target function within a gradient-free optimizer such as the downhill simplex \cite{preuhs2020} or the covariance matrix adaptation evolutionary strategy \cite{sisniega2017} to minimize it with respect to the geometry parameters. In contrast, we can rely on the geometry gradients and maximize the predicted SSIM of the trained network with respect to the rigid motion parameters $\alpha_i$, $t_{x,i}$, and $t_{y,i}$ for $i=1,...,N$ in a gradient-based manner. The target function is formulated as
\begin{equation}
    \label{eq:12}
    L = 1 - f_{\bm{\theta}}(\bm{I}) \enspace ,
\end{equation}
where $\bm{I} \in \mathbb{R}^{N_x\times N_y}$ is the reconstructed image. The gradient flow from the target function to the rigid geometry parameters relies on both \textit{PyTorch}'s automatic differentiation and the analytical gradients derived in Section~\ref{sec:gradients}: First, the gradient of the loss is backpropagated through the trained network with freezed weights into the reconstructed image using automatic differentiation. Second, gradient backpropagation from the reconstructed image to the updated projection matrices $P_i^*$ is performed with the analytical computations introduced in this paper. Finally, from the full updated projection matrices, the gradient is further backpropagated into just the rigid motion parameters $\alpha_i$, $t_{x,i}$, and $t_{y,i}$ by means of automatic differentiation.

\section{Experiments}

\subsection{Numerical Gradient Computation}
To prove the correctness of the analytical gradients, we compare them with a numerical approximation using forward differences with a small number $h > 0$. For simplicity, we perform these calculations on a contrast enhanced Shepp-Logan phantom with a target function computing the mean image intensity. 

\subsection{Data}
\label{sec:3.2}
We create a data set from motion-free reconstructed head CT scans which are publicly available under a TCIA Restricted License Agreement \cite{moen2021}. From each scan, we manually exclude slices which either contain mostly background or exhibit metal artifacts from, e.g., dentures or medical devices. Each slice is forward projected separately using $360$ projections on a full circle, a source to isocenter distance of \SI{1000}{\milli\meter}, and a source to detector distance of \SI{2000}{\milli\meter}. The detector has $1024$ elements of size \SI{2}{\milli\meter}. The images are of size $512 \times 512$ pixels and we assume an isotropic pixel spacing of \SI{1}{\milli\meter}. For artificial motion simulation, we sample four random motion patterns for each slice by first sampling an individual maximal motion amplitude $m_{max}$ for the rotations $\alpha_i$ and the translations $t_{x,i}$ and $t_{y,i}$. Then, a random perturbation is sampled for each of the three motion parameters by drawing $N$ independent samples from the uniform distribution $[-m_{max} / 2, m_{max} / 2]$. The maximum motion amplitudes present in the data set are \SI{4}{\milli\meter} for translation and $6.875^{\circ}$ for rotation. These random perturbations are added to the projection matrices as defined in Eq.~\ref{eq:11} and the simulated detector measurements are backprojected using the four perturbed as well as the one unperturbed set of projection matrices. This yields $5605$ samples from $50$ different patients. The SSIM between the motion-free and motion-corrupted reconstruction is computed as target metric. The data set is split on patient level into training (40 patients), validation (5 patients), and test set (5 patients). 

\subsection{Autofocus Network}
\label{sec:3.3}
A DNN is trained to predict the SSIM between motion-free and motion-corrupted reconstructed images given only the motion-corrupted image as input. The architecture consists of three randomly initialized convolutional layers with ReLU activation function followed by global average pooling. The loss function is the L1-distance between the predicted and the true SSIM value and the weights are optimized by an Adam optimizer with learning rate $\num{0.001}$ for $600$ epochs.

\subsection{Motion Compensation}
For the motion compensation experiments, slices from the test set are used for a fair estimation of performance. Again, a random perturbation is sampled for each projection and each of the three motion parameters $\alpha_i$, $t_{x,i}$, and $t_{y,i}$. We now use a fixed maximum motion amplitude of \SI{3}{\milli\meter} and $2.865^{\circ}$. The task is to recover the original, unperturbed projection matrices from the perturbed ones by finding rotation and translation parameters which optimally annihilate the initially added perturbation.
\begin{figure}[t!]
    \centering
    \includegraphics[width=\linewidth]{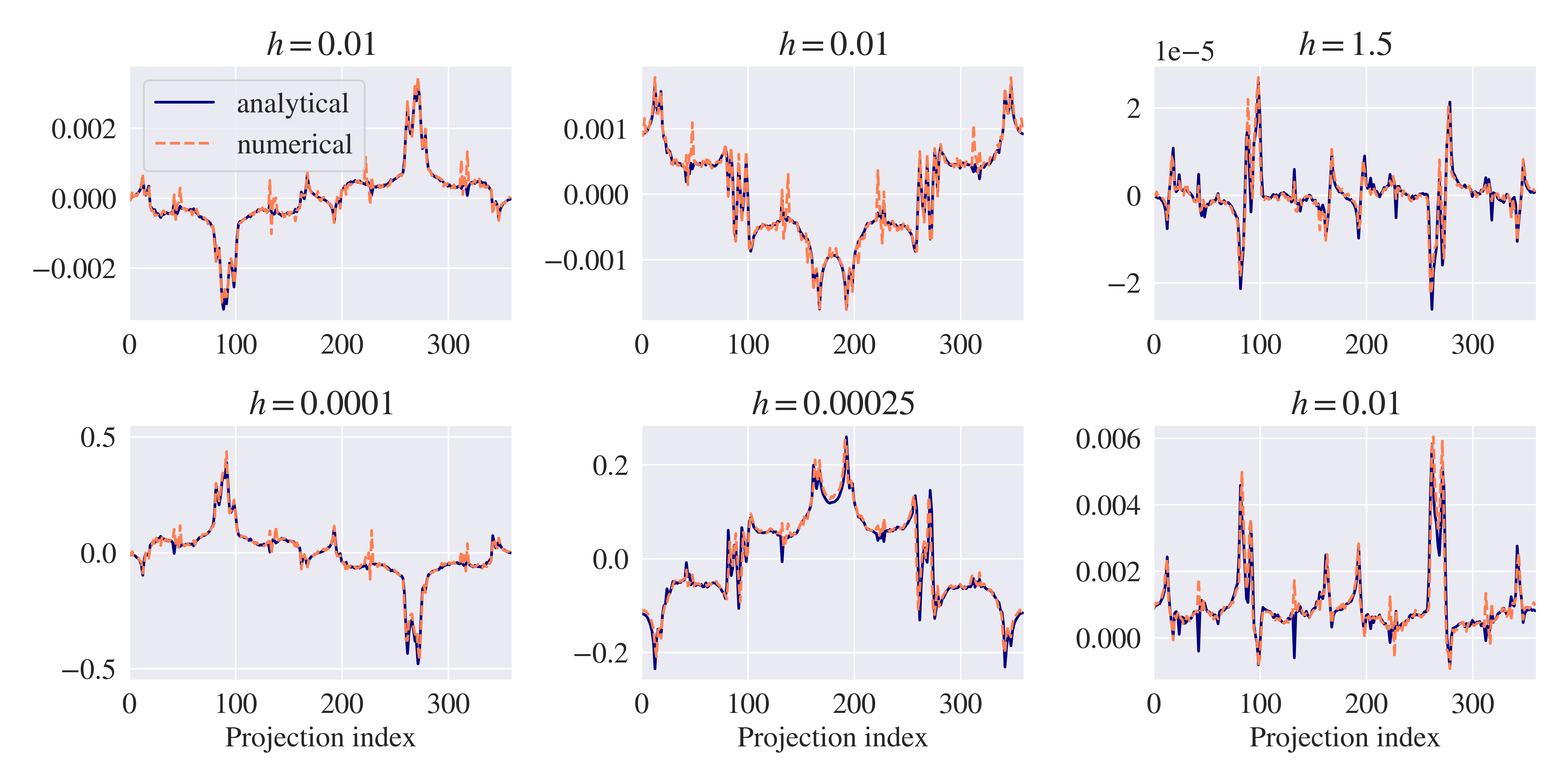}
    \caption{Comparison of analytical computation and numerical approximation of the gradient. The position of each subplot represents the position of the entry in the $2 \times 3$ projection matrices. Each curve visualizes the gradient of the mean image intensity with respect to this entry over all $360$ projections of the scan. The small perturbation $h$ used for the numerical gradient is indicated in the title of each subplot.}
    \label{fig:analytical_numerical}
\end{figure}
The differentiable motion compensation (1) applies the current guess of the annihilating motion parameters to the perturbed projections matrices following Eq.~\ref{eq:11}, (2) reconstructs the ramp filtered projection data from these projection matrices using the geometry-differentiable reconstruction operator (Section~\ref{sec:gradients}, Implementation), and (3) computes a loss based on the current reconstruction. This loss is minimized with respect to the annihilating motion parameters using gradient-based optimization. For the optimization, $500$ iterations of the stochastic gradient descent (SGD) algorithm without momentum implemented in \textit{PyTorch} are used with a step size of $0.1$ for the rotation parameters and a step size of $100$ for the translation parameters. To study the behavior of the implemented gradients, two different loss settings are investigated. First, the loss function is the mean-squared-error (MSE) between the current motion corrupted and the ground-truth reconstruction. This supervised loss is not a typical motion compensation setting as it assumes knowledge of a ground-truth, motion-free scan. Nevertheless, it can serve as an upper performance bound because the MSE loss yields a maximally informative image gradient which is subsequently propagated into the motion parameters using the gradient calculations proposed in this paper. As a more realistic setting which does not require a motion-free scan, we utilize the trained autofocus network (Section~\ref{sec:3.3}) with frozen weights and compute the loss function following Eq.~\ref{eq:12}. Consequently, in this setting, the SSIM predicted by the network based only on the motion corrupted reconstruction is maximized with respect to the free motion parameters using gradient descent.   

\section{Results}
We compare the analytical derivatives calculated in this paper with an approximation thereof using numerical differentiation. The results are depicted in Fig.~\ref{fig:analytical_numerical}. Each subplot contains the gradient with respect to one of the six entries in the projection matrices for all $360$ projections of the scan. Analytical and numerical estimates are almost identical. The value of the small number $h$ in the numerical computation which yields a precise approximation differs between the six entries of the projection matrices by several orders of magnitude. We manually select a suitable value $h$ which yields neither oversmooth (large $h$) nor instable (small $h$) approximations for each projection matrix entry separately.

The creation of the data set described in Section~\ref{sec:3.2} leaves us with a mean SSIM value of $0.83$. The range of SSIM values present in the data set spans from a minimal value of $0.56$ to a maximum of $0.97$. After training the network as described in Section~\ref{sec:3.3}, we achieve a mean absolute error of $\num{0.0056}$ concerning predicted SSIM on the held out test set.

\begin{figure}[t!]
    \centering
    \includegraphics[width=\linewidth]{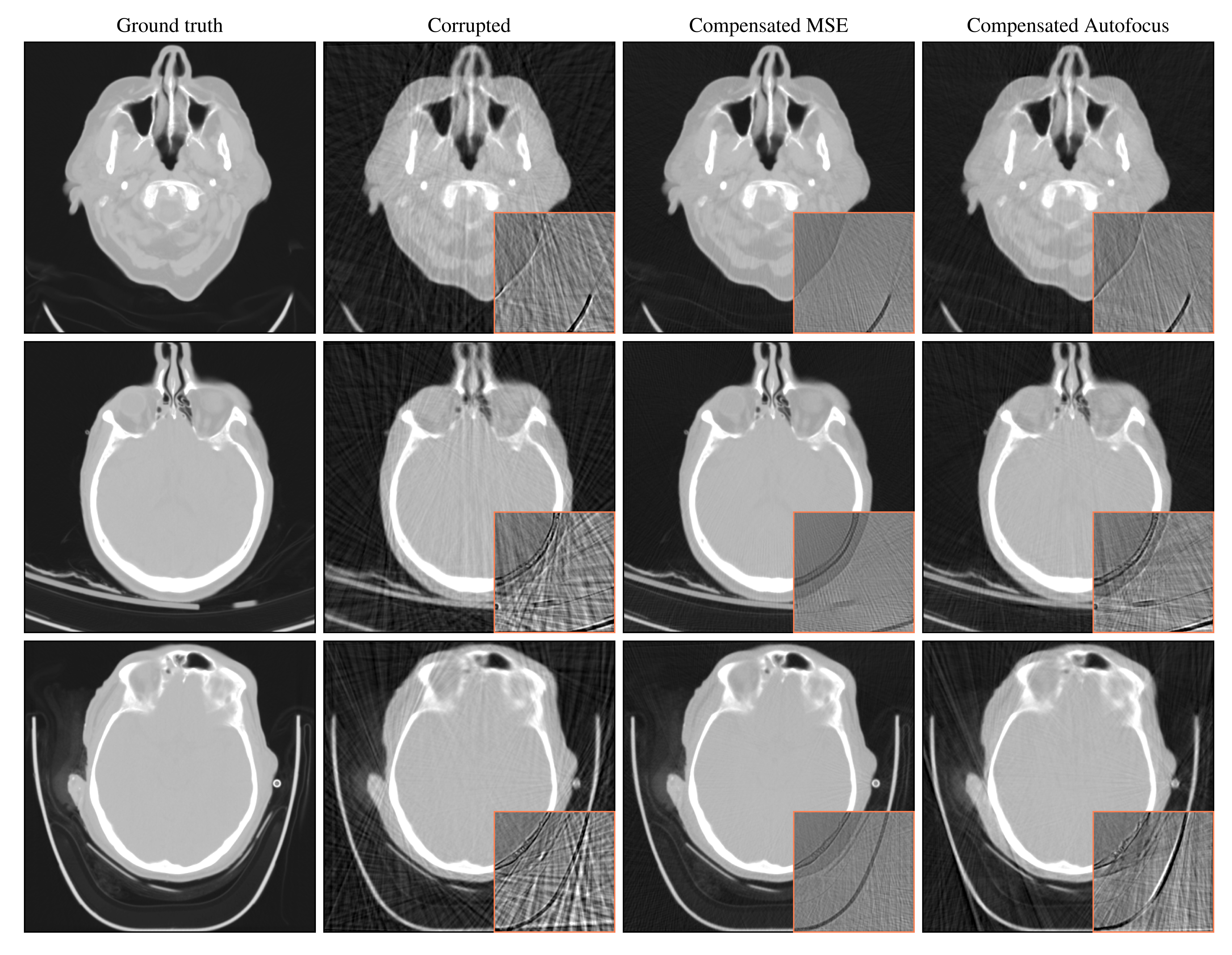}
    \caption{Qualitative reconstruction results for slices of three different patients. The columns contain the ground truth, the initial motion corrupted reconstruction, and the motion compensated reconstructions using MSE and autofocus objective. The orange frame contains the difference of that patch to the ground truth. The bottom row shows the slice with the \textit{worst} MSE after motion compensation using the autofocus objective. Gray values are windowed identically across all examples.}
    \label{fig:recos}
\end{figure}

Results of the gradient-based motion compensation are visualized in Fig.~\ref{fig:recos}. Whereas the motion corrupted reconstruction exhibits strong streaks and slightly blurred edges, both compensated reconstructions reduce these effects. In particular, the reconstruction compensated with MSE loss almost perfectly restores even fine details and low contrast structures such as the cushion in the background. The compensated reconstruction using the autofocus objective reduces streaks largely, but not entirely. Still, fine structures such as the nostrils appear much sharper and the background is more homogeneous compared to the uncorrected image. The bottom example in Fig.~\ref{fig:recos} shows the slice from the test set with the highest MSE to the ground truth after motion compensation with the autofocus objective. Nevertheless, fewer streaks are visible and the edges are sharper than the uncorrected version. Table~\ref{tab:quantitative} lists the quantitative results computed with respect to the ground-truth reconstruction. The compensation with MSE loss achieves the best results in terms of SSIM, MSE, and reprojection error (RPE). The compensation with the autofocus objective improves upon the uncorrected reconstruction with respect to all three metrics as well. For SSIM, the metric that the autofocus network is trained to predict, the method achieves an average of $0.89$ over all test images thereby improving upon the motion affected reconstruction by \SI{12.6}{\percent}.

\begin{table}[!t]
\setlength{\tabcolsep}{5pt}
\renewcommand{\arraystretch}{1.2}
\caption{Reconstruction quality of the motion corrupted and compensated images is quantified with respect to the ground-truth reconstruction.}
\label{tab:quantitative}
\centering
\begin{tabular}{ @{} l l l l l l l @{} }
\hline
                                    & SSIM      & MSE [$\cdot \num{1e-2}$]      & RPE [mm]       \\ 
\hline
Corrupted                    & $0.791 \pm 0.038$     & $5.717 \pm 1.340$    & $2.481 \pm 0.051$ \\  
Compensated MSE             & $0.965 \pm 0.008$     & $1.086 \pm 0.223$    & $0.649 \pm 0.224$ \\
Compensated Autofocus       & $0.891 \pm 0.024$     & $3.689 \pm 1.149$    & $2.276 \pm 0.392$ \\
\hline
\end{tabular}
\end{table}

\section{Discussion}
We showed how to derive fan-beam reconstructed CT images with respect to the matrices defining the acquisition geometry both mathematically and practically. We further demonstrated that our computations are correct. Such analytical differentiation is of interest in multiple ways: First, for optimization-based algorithms which formulate their target function on the reconstructed image but optimize geometry parameters, it opens the door to transition from gradient-free approaches to gradient-based optimizers. Second, it allows for gradient-flow into the geometry parameters in a deep learning sense thereby enabling loss functions on the reconstructed image which drive parameter updates in the geometry space. As an example, we propose a gradient-based rigid motion compensation algorithm which resides between the mentioned pure optimization algorithms on the one hand and full-blown deep learning approaches on the other hand. Using both \textit{PyTorch's} automatic differentiation and the proposed analytical derivatives, we optimize for the three rigid motion parameters for each projection of a fan-beam reconstruction. Because of this explicit problem formulation, we ensure that the updated matrices are valid projection matrices at any point. Optimization based on a MSE target function to the ground-truth motion-free image achieves motion compensated results which are very close to this ground truth. While this might seem obvious, it verifies that an informative gradient on the reconstructed image can be propagated into equally informative gradients on the geometry parameters. For a more realistic setting, we replace the MSE loss by an autofocus-inspired DNN trained offline to predict the SSIM to the ground-truth reconstruction given only the motion-corrupted reconstruction. Even with a very basic neural network architecture, this yields strongly improved results over the initial motion corrupted reconstructions. The autofocus network-based setting cannot be expected to perform competitively to the MSE loss setting because it has no access to the ground truth and, hence, the gradient with respect to the reconstructed image itself is less informative. However, the focus of this work is not the development of an ideal autofocus network, but to demonstrate how the gradient information obtained from such approach can be used for geometry optimization. Additionally, the random motion studied in our experiments is extremely hard to recover because it does not incorporate a smoothness constraint. Most related works on motion compensation assume smooth motion curves which reduces the degrees of freedom of the underlying optimization problem. Including more realistic motion models is subject to future work.      

The presented formulation derives a target metric with respect to the projection matrices because these are a very general and widely used concept for parameterization of CT acquisition geometry which can be extended to 3D cone-beam CT in a straight-forward manner. However, care must be taken when optimizing only for a specific subgroup of physical geometry parameters. For example, in the case of rigid motion compensation, only three extrinsic parameters are updated per projection while keeping the intrinsic parameters constant. Therefore, directly optimizing the six entries of the projection matrices would exceed the dimensionality of the problem. Instead, automatic differentiation can be leveraged to propagate the gradient from the full projection matrices into only the rigid motion parameters. Alternatively, the idea of Eq.~\ref{eq:6} could easily be extended by one more subgradient $\frac{\partial \bm{P}}{\partial \alpha} \rvert_{\alpha=\alpha_j}$. This modifies the analytical gradient itself to contain only, in this case, the information with respect to the rotation angle $\alpha$.
Finally, while this work focuses on an application to rigid patient motion compensation, a similar algorithm can be envisioned for phantom-free calibration of C-arm CT scanners \cite{aichert2018,maier2011}.

\section{Conclusion}
This work presents the calculations and implementations needed for gradient flow from the image domain into the geometry domain in fan-beam CT reconstruction. Thereby, it complements existing differentiable CT reconstruction operators which lack differentiability with respect to geometry parameters. Not only does this enable end-to-end training of deep models acting directly on the geometry parameters, but also the adoption of gradient-based strategies for optimization problems which have so far been limited to gradient-free optimizers. We envision a multitude of possible applications of the proposed method in gradient-based motion compensation and scanner calibration. Future work will also investigate the extension of the presented ideas to cone-beam geometries.
\enlargethispage{2\baselineskip}
\section*{Acknowledgements}
The research leading to these results has received funding from the European Research Council (ERC) under the European Union’s Horizon 2020 research and innovation program (ERC Grant No. 810316). The authors gratefully acknowledge the scientific support and HPC resources provided by the Erlangen National High Performance Computing Center (NHR@FAU) of the FAU Erlangen-N\"urnberg. The hardware is funded by the German Research Foundation (DFG). 

%
%
%
\bibliographystyle{splncs04}
\bibliography{bibliography}
\end{document}